\documentstyle[preprint,aps]{revtex}
\begin{document}

\tightenlines
\draft

\title{Dynamics and thermodynamics of a probe brane in the
  multicenter and rotating  D3-brane background}

\author{Rong-Gen Cai\footnote{email address: cai@het.phys.sci.osaka-u.ac.jp}}
\address{Department of Physics, Osaka University, Toyonaka,
      Osaka 560-0043, Japan}

\maketitle

\begin{abstract}
We study the dynamics and thermodynamics of a probe D3-brane in the 
rotating D3-brane background and in its extremal limit, which is a 
multicenter configuration  of D3-branes distributed uniformly on a disc.
In the extremal background, if the angular momentum of the probe does not 
vanish, the probe is always bounced back at some turning point. When its
angular momentum vanishes, in the disc plane, the probe will be captured
at the edge of the disc; in the hyperplane orthogonal to the disc, the
probe will be absorbed at the center of the disc. In the non-extremal
background, if the probe is in the hyperplane orthogonal to the disc, 
it will be captured at the horizon; if the probe is restricted in the 
disc plane,  the probe will be bounced back at a turning  point, which is 
just the infinite red-shift hyperplane of the rotating background,  even 
when the angular momentum of the probe vanishes.  The thermodynamics of a
relative static D3-brane probe is also investigated to the rotating D3-brane
source. Two critical points are found. One is just the thermodynamically 
stable boundary of the source rotating D3-branes; the other  is related to the
distance between the probe and the source, which can be regarded as the
mass scale in the corresponding super Yang-Mills theory. If the probe is 
static, the second critical point occurs as the probe is at the infinite 
red-shift hyperplane of the background. The relevance to the thermodynamics
of the super Yang-Mills theory is discussed briefly. 

\end{abstract}

\newpage

%%==================== section 1===============================

\section{Introduction}

Over the past several years the probe method has been used extensively in
investigating the structure of black holes, the bound state of branes, 
dynamics, thermodynamics and  statistical mechanics of branes, and so on  
\cite{Douglas,Mald1,Tseyt1,Chep1,Chep2,Liu,Youm,Kiritsis1,Kiritsis2,Land}. 
It turns out that many 
calculations involving the probe in  supergravity backgrounds are 
in  agreement with those obtained from the point of view of field theory.

On the other hand,  there are two branches in the  ${\cal N}$=4 super 
Yang-Mills (SYM) theory in four dimensions. The Higgs branch corresponds to 
 the vevs of scalar fields being  zero, while the gauge theory is in the
Coulomb branch when the vevs of some scalar fields do not vanish.
 According
to the Maldacena's conjecture \cite{Mald2}, different states in the
SYM theory can be described in terms of different configurations
in supergravity. For instance, a vacuum state of ${\cal N}$=4 SYM 
 theory with gauge group $U(N)$ in the Higgs branch is described 
by $N$ extremal D3-branes
coinciding with each other, and a thermal state of the theory is supposed
to be described by the near-extremal D3-branes. Furthermore, 
the states of the SYM  in the Coulomb branch are described
 by multicenter D3-brane solutions. A simplest case is that $N$ parallel 
 coinciding D3-branes are separated along a single transverse direction 
by a distance from a single D3-brane. In this case, the
 gauge symmetry $U(N+1)$ of the gauge theory is broken 
to $U(N)\times U(1)$.

 However, unlike the single-center  configurations, the multicenter 
solutions have not the non-extremal generalizations. This results
 in the difficulty to investigate thermodynamics of the SYM in Higgs phase. 
 Recently Tseytlin and Yankielowicz \cite{Tseyt2} attacked this issue and 
  studied the free energy of the  SYM in  the Higgs phase by using the 
 probe method. They interpreted the supergravity interaction potential
 between near-extremal D3-branes (as source) and a D3-brane (as a probe)
 as contribution of massive states to the free energy of the large N  
SYM  theory at strong 't Hooft coupling. In this method, it is worth noting 
that the source is excited, that is, the source is near-extremal static 
D3-branes, while the probe does not get excited. In this way some interesting 
results were observed. For example, from the free energy of the probe they 
predicted the existence of a phase transition. Note that there is no phase 
transition in the conformal case \cite{Witten}. Therefore phase transition may
 be present in the Higgs phase. In the low temperature limit they found that 
the structure of terms which appear in the free energy at strong and weak 
couplings is the same.

Recently, it has been found that it is possible to have the non-extremal 
generalizations of multicenter D3-brane configurations. But these D3-branes
are distributed continuously, rather than discretely. The
non-extremal generalizations of the continuously distributed D3-branes
are just the rotating D3-brane solution found in 
\cite{Russo,Kraus,Sfetsos,Russo1}, based on \cite{Cvetic}. It means that 
the SYM corresponding to the rotating D3-branes is in the Higgs phase. 
 Indeed, from
the calculations in supergravity, it is learned that the behavior of the 
SYM at strong 't Hooft coupling is quite different in the different branches.
For example, in the Higgs branch the quark-antiquark potential has the 
Coulombic behavior \cite{Mald3,Rey}, while in the multicenter backgrounds 
the potential has not only the Coulombic behavior, but also the confining 
behavior \cite{Mina,Freed,Brand}.Their thermodynamic behaviors are also quite 
different. 
The thermodynamics of the static D3-branes is stable; the heat capacity is
always positive-definite and hence the SYM is always in the unconfinement 
phase \cite{Witten}. Furthermore the so-called localization instability for
the static D3-brane configurations does not happen \cite{Peet}. For the 
rotating D3-branes, however, it has been found that the thermodynamics of 
excitations of D3-branes is stable up to a critical angular momentum 
density, beyond which the heat capacity will become negative, and a phase 
transition may happen \cite{Gubser1,Cai1,Gubser2}. The 
localization instability may occur when the angular momentum reaches 
some critical value \cite{Cai2}.

In this paper, we would like to investigate the structure and thermodynamics
of the multicenter and rotating D3-brane configurations by using the probe
method.  In the next section we analyze the dynamics of a D3-brane 
probe in this multicenter and rotating D3-brane background. The thermodynamics 
of the probe will be investigated in section III. A comparison to the case
of single-center solution is  also made. In section IV main results are 
summarized with a brief discussion.

%================ section 2 =====================================

\section{Dynamics of a probe D3-brane in the multicenter and
     rotating D3-brane background}

The D3-brane solutions in the type IIB supergravity have six transverse 
spatial dimensions. Therefore the  rotating D3-brane solutions may 
have three independent angular momentum parameters. For simplicity, we
 consider the rotating D3-brane solution with only an angular momentum 
parameter.  In this case, its metric is
\begin{eqnarray}
\label{e1}
ds^2 &=& \frac{1}{\sqrt{f}}\left(-h dt^2 +dx_1^2 +dx_2^2+dx_3^2 \right)
      +\sqrt{f}\left [\frac{dr^2}{\tilde{h}}-\frac{4ml\cosh \alpha}{r^4
            \triangle f}\sin^2\theta dtd\phi
	          \right. \nonumber \\
 &+& \left. r^2 (\triangle d\theta^2 +\tilde{\triangle}\sin^2\theta
              d\phi^2 +\cos^2\theta d\Omega_3^2) \right],
\end{eqnarray}
 where
\begin{eqnarray}
 && f=1+\frac{2m \sinh^2\alpha}{r^4\triangle},\ \ \
 \triangle =1+\frac{l^2\cos^2\theta}{r^2}, \ \ \
 \tilde{\triangle}=1+\frac{l^2}{r^2} +\frac{2m l^2\sin^2\theta}
       {r^6\triangle f}, \nonumber \\
&& h=1-\frac{2m}{r^4 \triangle}, \ \ \
 \tilde{h}=\frac{1}{\triangle}\left (1 +\frac{l^2}{r^2}-\frac{2m}{r^4}
   \right). 
\end{eqnarray}
In this solution the dilaton is a constant and the nonvanishing components of
the four-form potential are
\begin{equation}
\label{e2}
C=-\frac{(f^{-1}-1)}{\sinh\alpha}dx_1\wedge dx_2\wedge dx_3 \wedge
     (\cosh\alpha dt -l \sin^2\theta d\phi).
\end{equation}
Corresponding to this rotating D3-brane solution, the R-symmetry of SYM theory
is broken from $SO(6)$ to $SO(4)\times U(1)$. Furthermore, it is observed that 
 the extremal limit of this
solution, $m \rightarrow 0$ keeping $l$ fixed and $m e^{2\alpha}$ finite, is
a multicenter solution, which represents D3-branes to be distributed 
uniformly on a disc with a radius $l$\cite{Kraus,Sfetsos}. That is, 
 this extremal limit is some superposition of 
$N$ static D3-branes, rather than that of $N$ coinciding rotating D3-branes.
Taking an appropriate coordinate transformation, this limiting metric becomes
\begin{equation}
\label{e3}
ds^2=H_0^{-1/2}(-dt^2 +dx_1^2 +dx_2^2 +dx_3^2) +H_0^{1/2}(dy_1^2 +dy_2^2
       +dy_3^2 +dy_4^2 +dy_5^2 +dy^2_6),
\end{equation}
where 
\begin{equation}
H_0= 1+\frac{2R^4}{[r^2-l^2 +\sqrt{(r^2+l^2)^2-4l^2\rho ^2}]
          \sqrt{(r^2+l^2)^2 -4l^2\rho ^2}}.
\end{equation}
Here $r^2 =y_1^2 +\cdots + y_6^2$, $\rho^2 =y_5^2 +y_6^2$, and $R^4
=4\pi g_sN\alpha'^2$. The four-form 
potential reduces to
\begin{equation}
\label{e5}
C=(H_0^{-1}-1)dt\wedge dx_1\wedge dx_2 \wedge dx_3.
\end{equation}

On the other hand, the dynamics of  a D3-brane probe is governed 
 by the following action:
\begin{equation}
\label{e6}
S=-T_3\int d^4x \sqrt{-{\rm det}\, \hat{G} } -T_3\int \hat{C},
\end{equation}
where $T_3=1/2\pi g_s (2\pi\alpha')^2 =N/2\pi^2 R^4$ is the D3-brane tension.
In order to investigate the dynamics of the probe, it is convenient to 
take the static gauge: $\tau=t$, $x_i$ act as just the worldvolume coordinates,
and other transverse coordinates $y_i$ depend on $\tau$ only.
Let's first discuss the case of the probe in the extremal background
(\ref{e3}). 

%%==================subsection 1========================

\subsection { In the extremal background}

Substituting (\ref{e3}) and (\ref{e5}) into (\ref{e6}),
we obtain 
\begin{equation}
\label{e7}
S=-T_3V_3 \int d\tau H_0^{-1}[\sqrt{1-H_0\dot{y}_i^2}-1],
\end{equation}
where  the overdot denotes derivative with respect to $\tau$, 
$V_3$ is the spatial volume of the worldvolume and we subtracted a 
constant term. For a static probe, it is obvious  from (\ref{e7}) 
that the interaction potential vanishes, which  means that the source is 
indeed a BPS configuration.  For a low-velocity probe, its action is,
up to the term ${\cal O}(v^4)$,
\begin{equation}
S=\frac{T_3V_3 }{2}\int d\tau \dot{y}_i^2 + {\cal O}(v^4).
\end{equation}
It is seen  that the motion of the probe is like  the   motion of a test 
particle with mass $m_p=T_3V_3$ in a flat spacetime. Therefore,  to this order,
its motion is the same as that in a single-center D-brane background 
\cite{Mald1}. Of course, this is also required by the BPS property of 
the system consisting of the source and the probe.

    Now we consider a general motion of the probe in the extremal background.
Generally  speaking, the motion of the probe is like that of  a test particle 
with mass $m_p$ moving in a velocity-dependent potential. Set the probe
  to have the angular momentum 
only in some $\phi$-direction, we can then write down: $\dot{y}_i^2=\dot{r}^2 
+r^2 \dot{\phi}^2$. From (\ref{e7}) we have the angular momentum $L$ of 
the probe, 
\begin{equation}
  L=\frac{m_pr^2 \dot{\phi}}{\sqrt{1-H_0(\dot{r}^2 +r^2\dot{\phi}^2)}}.
\end{equation}
And the energy $E$ of the probe is
\begin{equation}
E=\frac{m_p(\dot{r}^2 +r^2\dot{\phi}^2)}{\sqrt{1-H_0(\dot{r}^2
        +r^2\dot{\phi}^2)}}+ m_pH_0^{-1}[\sqrt{1-H_0(\dot{r}^2
	+r^2\dot{\phi}^2)}-1].
\end{equation}
From the above equation, using the expression of the angular momentum and 
the kinetic relation,
\begin{equation}
E=\frac{1}{2}m_p\dot{r}^2 +V(r),
\end{equation}
 we can obtain an effective potential of the radial motion of the probe,
\begin{equation}
V(r)=E\left [1-\frac{1+EH_0/2m_p}{(1+EH_0/m_p)^2}\right]
        +\frac{L^2}{2m_p r^2 }\frac{1}{(1+EH_0/m_p)^2}.
\end{equation}
Thus, the radial motion of the probe is that of a test particle with 
mass $m_p$ moving in the velocity-independent central 
force potential $V(r)$. Characterizing 
the motion is the turning points, which satisfy the equation
 $E=V(r_c)$. It can be seen clearly
from the effective potential that when the angular momentum vanishes, $L=0$,
there is no turning point, and then the probe might  be captured by
the source.

 As mentioned above, the source D3-branes are distributed on a disc in a 
 plane defined by $y_1=y_2=y_3=y_4=0$. So it would be interesting to study
 the motion of the probe in this plane or orthogonal to this plane. Let's 
first discuss the case for the probe moving in the disc plane.

(i).{\it  In the disc plane}. In this case, 
one has  $r ^2=\rho^2$, and
 \begin{equation}
\label{e13}
 H_0= 1 +\frac{R^4}{(\rho^2-l^2)^2},
 \end{equation}
where $\rho^2 =y_5^2 +y_6^2$.
 Indeed, the harmonic function is singular at the edge of the disc $\rho=l$. 
 In this case, the effective potential becomes
 \begin{equation}
 V(\rho)=E\left [1 - \frac{1 +\frac{E}{2m_p}\left (1+
       \frac{R^4}{(\rho^2-l^2)^2}
        \right)}{\left[1+\frac{E}{m_p}\left (1 +\frac{R^4}{(\rho^2-l^2)^2}
	\right) \right ]^2} \right ] 
	+ \frac{L^2}{2m_p\rho ^2}\frac{1}{\left [1+\frac{E}{m_p}
	\left (1 +\frac{R^4}{(\rho^2-l^2)^2}\right)\right]^2 }.
 \end{equation}
We first consider the far region, that is, $\rho ^2-l^2 >>R^2$. The potential 
has the following behavior:
\begin{eqnarray}
V(\rho) \approx && E\left [1 -\frac{m_p(E+2m_p)}{2(E+m_p)^2}
     +\frac{m_pE(E+3m_p)}{2(E+m_p)^3}\frac{R^4}{(\rho^2-l^2)^2}\right]
      \nonumber \\
      &+& \frac{m_pL^2}{2(E+m_p)^2\rho^2}
      \left [1-\frac{2E^2}{m_p (E+m_p)}\frac{R^4}{(\rho^2-l^2)^2}\right].
\end{eqnarray}
The force exerted  on the probe mainly comes from the repulsive centrifugal
force due to the non-zero angular momentum $L$. The effect  
 of the source is of the  sub-leading order. Compared to the single-center case,
where $l=0$, the sub-leading  effect  is enhanced due to the distribution 
of the source D3-branes.  In  the region near the edge of the disc, namely, 
$\rho^2-l^2<< R^2$,  the constant $1$ in 
the harmonic function (\ref{e13}) can be dropped out. The effective potential 
reduces to
\begin{equation}
 V(\rho)=E\left [1 - \frac{1 +\frac{\rho_* ^4}{2 (\rho^2-l^2)^2}}
        {\left (1+\frac{\rho_*^4}{(\rho^2-l^2)^2}
	 \right )^2} \right ] 
	+ \frac{L^2}{2m_p\rho ^2}\frac{1}{\left (1+\frac{\rho_*^4}
	{(\rho^2-l^2)^2}\right)^2 },
 \end{equation}
where we introduced a characteristic scale $\rho_*^4 =E R^4/m_p$.
At the distance $\rho^2-l^2 >> \rho_*^2$, the effective potential takes
the following form:
\begin{equation}
V(\rho) \approx \frac{3E\rho_*^4}{2(\rho^2-l^2)^2}+\frac{L^2}{2m_p\rho^2}
     \left (1-\frac{2 \rho_*^4}{(\rho^2-l^2)^2}\right ).
\end{equation}
In this region, the potential is a sum of two repulsive potentials. 
The first term  is enhanced due to
the distribution of the source, while the usually  repulsive centrifugal 
potential (second term) is suppressed. 
 When $\rho^2 -l^2 <<\rho^2_*$, 
\begin{equation}
\label{e18}
V(\rho) \approx E -\frac{E (\rho^2-l^2)^2}{2\rho_*^4} +\frac{L^2}{2m_p\rho^2}
      \frac{(\rho^2-l^2)^4}{\rho_*^8}.
\end{equation}
The repulsive  potential (second term)  in fact is independent of the energy
and is suppressed in this case; 
the usually centrifugal  potential (third term) becomes attractive and it 
is also suppressed due to the distribution effect of the source. 
 From (\ref{e18}) we see that the potential is
a constant $E$ at the edge of the disc and  the central force, $F(\rho) 
=-d V(\rho)/d\rho$, on the probe 
vanishes there. When $L^2=0$, the turning point is at $\rho^2_c =l^2$, 
 therefore, the probe will be captured and it will locate at $\rho_c=l$;
 when $L^2 \ne 0$, the turning point $\rho_c$ is
 \begin{equation}
\rho_c =\frac{\rho_*^2/\rho_{**} +\sqrt{(\rho_*^2/\rho_{**})^2 
      +4l^2}} {2},
\end{equation}
where $\rho_{**}^2=L^2/m_pE $. Therefore  we can conclude the motion of the
probe in the disc plane as follows. When the probe is far from the disc, the 
effect due to the distribution of source is of the  sub-leading order. For the 
radial motion in the disc plane, the probe will be captured at the edge of the 
disc if its angular momentum vanishes; the probe will be bounced back and 
never be absorbed by the source if its angular momentum does not vanish.

(ii). {\it In the hyperplane orthogonal to the disc}.
 When the probe is restricted in a hyperplane defined as $y_5=y_6=0$, which
is orthogonal to the disc plane discussed above, we have $\rho=0$, and
 \begin{equation}
\label{e20}
 H_0=1 +\frac{R^4}{r^2 (r^2+l^2)},
 \end{equation}
where $r^2=y_1^2 +y_2^2 +y_3^2 +y_4^2$.
In this case, $r=0$ is a singular hyperplane for the harmonic function.
 In the asymptotically
far region, we have the effective potential 
\begin{eqnarray}
V(r)\approx && E\left[1-\frac{m_p(E+2m_p)}{2(E+m_p)^2}
   + \frac{m_pE(E+3m_p)}{2(E+m_p)^3}\frac{R^4}{r^2(r^2+l^2)}\right]
      \nonumber \\
      &+&\frac{m_pL^2}{2(E+m_p)^2r^2}\left [1-\frac{2E^2}{m_p(E+m_p)}
      \frac{R^4}{r^2(r^2+l^2)}\right].
\end{eqnarray}
Once again, as expected, the distribution effect of the source is
in the sub-leading order and the main contribution to the  force on the probe  
comes from the repulsive centrifugal force. Unlike the case in the disc plane, 
however, the sub-leading  effect is suppressed here. 
In the region near the disc,  the effective potential
has the following form: 
\begin{equation}
 V(r)=E\left [1 - \frac{1 +\frac{\rho_* ^4}{2r^2 (r^2+l^2)}}
        {\left (1+\frac{\rho_*^4}{r^2(r^2+l^2)}
	 \right )^2} \right ] 
	+ \frac{L^2}{2m_p r^2}\frac{1}{\left (1+\frac{\rho_*^4}
	{r^2 (r^2+l^2)}\right)^2 }.
 \end{equation}
At the distance $r >> \rho_*$, the effective potential takes the following
form:
\begin{equation}
V(r) \approx \frac{3E\rho_*^4}{2r^2(r^2 +l^2)}+\frac{L^2}{2m_p r^2}
    \left(1- \frac{2\rho_*^4}{r^2(r^2+l^2)}\right ).
\end{equation}
Here the first  repulsive  potential term is suppressed, while the centrifugal 
repulsive potential is enhanced. When $r <<\rho_* $, the effective potential
 reduces to
\begin{equation}
\label{e24}
V(r) \approx E -\frac{E r^2(r^2+l^2)}{2\rho_*^4}
     +\frac{L^2}{2m_p } \frac{r^2 (r^2+l^2)^2}{\rho_*^8}.
\end{equation}
In this case, once again the usually centrifugal repulsive potential becomes
attractive and it is enhanced,
 compared to the single-center case. The second term in (\ref{e24}) is a 
repulsive potential, which is also enhanced due to the distribution effect
of the source. From (\ref{e24}) we see that 
the potential  is a constant $E$ at $r=0$ and  the central force
on the probe vanishes there. Therefore the probe will be absorbed at $r=0$
if the probe has no angular momentum. Otherwise, the probe will
be bounced back at the turning point:
\begin{equation}
r_c^2=\frac{\rho_*^4}{\rho_{**}^2}-l^2.
\end{equation}

%%========== subsection 2 ========================================

\subsection{ In the non-extremal background}

 In this case, the background
represents the rotating D3-brane solution. Substituting (\ref{e1}) 
and (\ref{e2}) into (\ref{e6}) yields
\begin{equation}
S=-T_3V_3\int d\tau f^{-1}\left [\sqrt{h-f\omega ^2}-1 +f_0 -f +
   \frac{(1-f)l\sin^2\theta}{\sinh\alpha}\dot{\phi}\right],
\end{equation}   
where 
\begin{equation}
\omega ^2= \frac{\dot{r}^2}{\tilde{h}} +r^2( \triangle \dot{\theta}^2
     +\tilde{\triangle}\sin^2\theta \dot{\phi}^2 +\cos^2\theta
     \dot{\Omega}_3^2) -\frac{4m l\cosh \alpha}{r^4 \triangle f}
     \sin^2\theta \dot{\phi}.
\end{equation}
and
\begin{equation}
f_0=1 +\frac{R^4}{r^4\triangle}.
\end{equation}
In this case, the static interaction potential will no longer vanish and 
can be written down as:
\begin{equation}
V_0(r,\theta)=m_p f^{-1}(\sqrt{h}-1+f_0 -f).
\end{equation}
The potential depends on not only the radial position of the probe, but also 
the azimuth $\theta$.  Quite interesting is that when 
$\theta =\pi/2$, which means that the probe is in the disc plane discussed 
above, the effect due to the rotation  of the source disappears. 
 The form of potential implies that the
static probe suffers from an attractive force from the source. The general
motion of the probe is quite complicated. Here 
we consider the motion of probe in two self-consistent cases: One
is $\theta=0$ and $\Omega_3=const.$, and 
the other  is $\theta=\pi/2$ and $\Omega_3=const.$.
In the first case, which implies that the probe is moving in the 
 hyperplane orthogonal to the disc plane,  the action reduces to 
\begin{equation}
\label{e30}
 S=-m_p\int d\tau f^{-1}[\sqrt{h-f\dot{r}^2/\tilde{h}}-1+f_0-f],
\end{equation}    
and
\begin{equation}
f=1+\frac{\tilde{R}^4}{r^4 \triangle}, \ \ \triangle=\tilde{\triangle}=
    1+\frac{l^2}{r^2}, \ \ h=1-\frac{2m}{r^4 \triangle}=
 \tilde{h}=\frac{1}{\triangle}\left (1+\frac{l^2}{r^2}-\frac{2m}{r^4}
     \right),
\end{equation}    
where $\tilde{R}^4=\sqrt{R^8+m^2}-m$. The dependence of  motion 
on the angular coordinate $\phi$ disappears automatically,  
the radial effective potential can be expressed as 
\begin{equation}
V(r)=E\left [1-\frac{m_p h\tilde{h}}{2Ef}\left(1-\frac{h}
     {(1+f-f_0+Ef/m_p)^2} \right)\right].   
\end{equation}
We are interested in the so-called  field theory limit, in which 
one has $f \approx f_0 = R^4/r^4\triangle$. And then the effective 
potential reduces to
\begin{equation}
\label{e33}
V(r)=E\left [1-\frac{r^4 \triangle h \tilde{h}}{2 r_*^4}
   \left(1-\frac{h}{(1+r_*^4/r^4\triangle)^2}\right)\right].
\end{equation}                                         
Here $r_*^4= ER^4/m_p$. For the rotating D3-brane solution (\ref{e1}), 
there is a horizon $r_+$, determined by $\tilde{h}=0$,
\begin{equation}
\label{horizon}
r_+^2=\frac{1}{2}\left(\sqrt{l^4+8m}-l^2\right).
\end{equation}
From (\ref{e33}) we find that the potential is a constant $E$ at the horizon
$r_+$ and the central force, $F(r)=-dV(r)/dr$, on the probe vanishes there. 
Furthermore, the turning point is just the horizon.  Therefore the
 probe will be absorbed by the source and will locate at the horizon.
Note that in this case the infinite red-shift hyperplane and event horizon
of the background coincide with each other.

In the second case, namely, $\theta=\pi/2$ and $\Omega_3=const.$,
 due to the dependence on the angular coordinate $\phi$, 
the motion is still complicated. To simplify the problem, we further set
$\dot{\phi}=0$, namely, the angular momentum of the probe 
vanishes in this simplified 
situation.  Thus the action  of the probe is also expressed by  (\ref{e30}),
but 
\begin{equation}
f=1+\frac{\tilde{R}^4}{r^4},\ \  f_0= 1 +\frac{R^4}{r^4}, 
     \ \ h=1-\frac{2m}{r^4}, \ \
 \tilde{h}=1+\frac{l^2}{r^2}-\frac{2m}{r^4}. 
\end{equation}    
In the field theory limit, the effective potential becomes
\begin{equation}
\label{e36}
V(r)=E\left [1-\frac{r^4  h \tilde{h}}{2 r_*^4}
   \left(1-\frac{h}{(1+r_*^4/r^4)^2}\right)\right].
\end{equation}                                         
In this case we have two turning points: one is the horizon $r_+$ and
the other  is $r_c^4=2m$, determined by $h=0$. Note that the latter
is just the infinite red-shift hyperplane in this case.
 At these two turning points 
the potential is the constant $E$, but the central force does not vanish there.
Because of $r_c>r_+$ the probe will be bounced back at $r_c$. 
To compare this to 
the case of static source is interesting. For the static source case, one has 
$h=\tilde{h}=1-2m/r^4$. In that case, the turning point coincides with 
the horizon and the force on the probe vanishes. Thus the probe will be 
captured and will locate at the horizon.

In the calculations of the Wilsonian potential in the non-extremal 
supergravity solutions (for example see \cite{Brand2,Rey2}), it has been 
assumed that the non-extremal branes ``locate'' at the horizon of
non-extremal backgrounds. The analysis in this section provides 
an evidence of this assumption, at least for static background: indeed
the brane can locate at the horizon of backgrounds. However, for the 
rotating background, the dynamics of the brane restricted in the disc plane
poses a puzzle: the vanishing angular momentum brane cannot locate at the 
horizon. Perhaps we should consider the probe brane with the same angular
velocity with the background. For such a brane, we expect that it can locate
at the horizon.

%%%%%%%%%%%%%%%%%% section 3     %%%%%%%%%%%%%%%%%%%%%%%%%%%%%

\section{Thermodynamics of a probe D3-brane in the rotating D3-brane 
         background}

Near the extremal limit, some thermodynamic quantities of the rotating D3-brane
solution (\ref{e1}) can be expressed as follows\cite{Cai1}:
\begin{eqnarray}
&& E=3 \pi ^3 \kappa^{-2} m V_3,  \nonumber \\
&& J= \pi ^{7/4}\kappa^{-3/2}N^{1/2}m^{1/2}lV_3, \nonumber  \\
&& \Omega =\pi^{5/4}\kappa^{-1/2}N^{-1/2}m^{-1/2}l r_+^2, \nonumber  \\
&& T=2^{-1}\pi^{1/4}\kappa^{-1/2}N^{-1/2}m^{-1/2}(2r^3_+ +l^2r_+),
 \nonumber \\
 \label{thermq}
 &&  S=2 \pi^{11/4}\kappa^{-3/2}N^{1/2}m^{1/2}r_+V_3.
 \end{eqnarray}
 Here $r_+$ is given by (\ref{horizon}); $E$ denotes the energy above the 
 extremality which equals the
 ADM mass of the black three-brane minus the mass of the
 corresponding extremal one; and $J$, $\Omega$, $T$ and $S$
 represent the angular momentum, angular velocity, Hawking temperature,
 and the entropy, respectively. In addition, $2\kappa^2 =16\pi G
 =(2\pi)^7 g_s^2 \alpha'^4$ and $G$ is the Newton gravitational constant in ten 
 dimensions. These quantities satisfy the first
 law of thermodynamics
 \begin{equation}
 \label{first}
 dE=TdS +\Omega dJ.
 \end{equation} 
In \cite{Cai1} it has been observed   that the heat capacity at a 
constant angular velocity, $C_{\Omega}=T(\partial S/\partial T)_{\Omega}$, 
diverges at $l^2=2r_+^2$, namely, $l^4/m =8/3$, and beyond which the heat 
capacity becomes negative. 
This means that the ${\cal N}$=4 large $N$ SYM corresponding to the rotating 
D3-brane configuration has a phase transition. Its critical behavior at the 
critical point  has been investigated and some relevant critical exponents 
have been obtained  in \cite{Cai1}.  Recall that the SYM theory corresponding
to this rotating D3-brane solution is in the Higgs phase. Thus the occurrence 
of phase transition in this rotating configuration is in agreement with the 
observation that phase transitions may be present in the Higgs phase 
\cite{Tseyt2}.

Corresponding to this thermodynamic system,  
the Gibbs free energy, which is defined as $G=E-TS-\Omega J$, is
\begin{equation}
\label{G_N}
G_N=-\frac{1}{2}\pi ^3 \kappa^{-2 }r_+^2 (r_+^2 +l^2)V_3,
\end{equation}
and the Helmholtz free energy, defined as $F=E-TS$, is
\begin{equation}
\label{F_N}
F_N=-\frac{1}{2}\pi^3 \kappa^{-2} r_+^2 (r_+^2-l^2)V_3.
\end{equation}
The Gibbs free energy does not change its sign, but Helmholtz free energy 
does at $r_+^2=l^2$, that is, $l^4/m=1$.  However, it is  not clear
by now whether  this change of sign is related or not to the Hawking-Page 
phase transition, which  takes place in Schwarzschild-anti-de Sitter 
black holes. This phase transition has been interpreted  by Witten as the 
confinement/unconfinement phase transition in the SYM theory \cite{Witten}.

Next we consider the thermodynamics of a probe D3-brane.  
 For a static probe, as explained by 
Tseytlin and Yankielowicz \cite{Tseyt2}, its distance to the source can be 
regarded
as a mass scale in the SYM, and hence interaction static potential between the 
source and the probe can be interpreted as a contribution of massive states to
the free energy of the large $N$ SYM at the strong 't Hooft coupling. In the 
rotating D3-brane background (\ref{e1}), to keep relative static position of 
the probe to the rotating source, we have to let the probe rotate along an 
orbit in the direction $\phi$ with the same angular velocity $\dot{\phi}
=\Omega$. The Euclidean action of such a probe is
\begin{eqnarray}
\label{action}
I\equiv \beta G_p &=& T_3V_3 \beta f^{-1}  
        \left( \sqrt{h-fr^2\tilde{\triangle} 
        \Omega^2 \sin^2\theta +\frac{4ml\Omega \cosh\alpha}{r^4\triangle }
  \sin^2\theta }   \right. \nonumber \\
  &-& \left. 1+ f_0-f +\frac{(1-f)l\Omega }{\sinh\alpha}\sin^2\theta 
   \right),
\end{eqnarray}
where $\beta=1/T$ is the inverse Hawking temperature. Since we are interested
in thermodynamics of SYM theory through the Maldacena conjecture, we consider
the field theory limit:
\begin{equation}
\label{limit}
r \rightarrow u \alpha', \ \ m \rightarrow m\alpha'^4, \ \ 
    l \rightarrow l\alpha',\ \  \alpha' \rightarrow 0,
\end{equation}    
but $g_s$ gets fixed. In this limit, we have $r_+^2 \rightarrow r_+^2\alpha'^2$
and all quantities in (\ref{thermq}) keep same forms, only difference is that
$2\kappa^2$ is replaced by $2\tilde{\kappa}^2 =(2\pi)^7g_s^2$. Furthermore,
equations (\ref{first}), (\ref{G_N}) and (\ref{F_N}) remain same forms as well.

Considering the field theory limit (\ref{limit}), we have 
\begin{equation}
f\approx f_0=\frac{R^4}{\alpha'^4 \triangle u^4}, \ \
        \triangle =1 +\frac{l^2 \cos^2\theta}{u^2}, \ \ 
     \tilde{\triangle}=1+\frac{l^2}{u^2} , \ \ 
     h=1-\frac{2m}{u^4\triangle}.
\end{equation}
And the free energy of the probe reduces to
\begin{eqnarray}
G_p &=& \frac{T_3V_3\alpha'^4\triangle u^4}{R^4}\left (\sqrt{1-\frac{2m}
     {u^4 \triangle}-\frac{R^4 \tilde{\triangle}\Omega^2}{\alpha'^2 \triangle 
     u^2}\sin^2\theta +\frac{2\sqrt{2m}l\Omega R^2}{\alpha' \triangle u^4}
     \sin^2 \theta }  \right.  \nonumber \\
     &-& \left. 1 -\frac{\sqrt{2m}l\Omega R^2}{\alpha' \triangle u^4}
     \sin^2\theta \right).
\end{eqnarray}
Having considered $\alpha'$ appearing in the tension $T_3$  and $R^4$,
 the parameter $\alpha'$ disappears in the above free energy, which makes 
 the contribution of the massive states  to the free energy of the SYM 
 theory be finite, 
\begin{equation}
\label{free}
G_p=\frac{V_3N\triangle u^4}{2\pi^2 \lambda ^2}\left ( \sqrt{1-\frac{2m}
     {u^4 \triangle}-\frac{\lambda \tilde{\triangle} \Omega^2}{\triangle u^2}
     \sin^2\theta +\frac{2\sqrt{2m\lambda}l\Omega}{\triangle u^4}
     \sin^2 \theta } 
     -1 -\frac{\sqrt{2m\lambda}l\Omega}{\triangle u^4}\sin^2
     \theta \right),
\end{equation}
where $\lambda =2Ng_{\rm YM}^2 =4\pi Ng_s$ is the 't Hooft coupling constant.
Unlike the case of  the probe in the static non-extremal D3-brane 
background, there the free energy of the probe depends on the temperature of
source and the mass scale $u$ of scalar fields (see \cite{Tseyt2}), here 
the free energy depends not only on the temperature and mass scale, but
also on the angular velocity of the source and the azimuth $\theta$.
In fact,  the 
dependence on the azimuth $\theta$ means that the probe is separated by not
only a single direction from the source D3-branes. In principle, 
we can eliminate 
$m$ and $l$ from (\ref{free}) by using the expressions of the Hawking
temperature and angular velocity in (\ref{thermq}). In practice, however, 
the expression of the free energy would be quite complicated if we did. 
So we keep the form (\ref{free}) in the following discussions.

From the expression (\ref{free}) of the free energy we see that there is a
minimum, which occurs when the square root in (\ref{free}) is zero, namely,
\begin{equation}
\label{crit}
     1-\frac{2m} {u^4 \triangle}-\frac{\lambda \tilde{\triangle} 
     \Omega^2}{\triangle u^2}
     \sin^2\theta +\frac{2\sqrt{2m\lambda}l\Omega}{\triangle u^4}
     \sin^2 \theta =0. 
\end{equation}
Having been given the free energy of the probe, we can acquire other
thermodynamic quantities of the probe immediately. For example, its entropy,
angular momentum, 
and heat capacity can be obtained by using following formulas: 
\begin{equation}
S_p=-\left(\frac{\partial G_p}{\partial T}\right)_{\Omega},
        \ \ \ \  
J_p=-\left(\frac{\partial G_p}{\partial \Omega}\right)_{T},
          \ \ \ \
     C_{\Omega p}=T\left(\frac{\partial S_p}{\partial T}\right)_{\Omega}.
\end{equation}

Below we  discuss two special positions of the probe.
One is  $\theta =0$.  That is, the probe is in a hyperplane 
perpendicular to the disc plane of the source.
The free energy (\ref{free})  takes  a very simple form
in this case:
\begin{equation}
\label{perp}
G_{\perp}=\frac{V_3 N u^2 (u^2 +l^2)}{2\pi^2 \lambda^2}
     \left (\sqrt{1-\frac{2m}{u^2(u^2+l^2)}}-1\right).
\end{equation}
The entropy of the probe is
\begin{equation}
S_p = \frac{2 V_3N\sqrt{2m\lambda}l^4}{\pi \lambda^2 (2r_+^4 
     +3l^2r_+^2 +l^4)}
   \frac{l^2+2r_+^2}{l^2-2r_+^2}
   \frac{1}{\triangle_*}
   \left [\frac{m (l^2-2r_+^2)}{l^4}
 - \frac{u^2(u^2+l^2)-m}{u^2+l^2} -u^2\triangle_*
   \right],
\end{equation}   
and its angular momentum 
\begin{equation}
J_p = \frac{V_3N}{\pi^2\lambda^2}
        \frac{\sqrt{2m\lambda}(2r_+^2+l^2)}{(r_+^2+l^2)(l^2-2r_+^2)}
	\frac{l}{\triangle_*} 
     \left [\frac{m(l^2-2r_+^2)}{r_+^2(2r_+^2+3l^2)}
      +\frac{u^2(u^2+l^2)-m}{u^2+l^2}
      -u^2\triangle_*\right]
\end{equation}      
where $\triangle _*=\sqrt{1-2m/u^2(u^2+l^2)}$.  We can also calculate the heat
capacity of the probe, but because of the complexity of its expression, 
we are not going to present it here.  We find that the entropy and angular 
momentum of the probe diverge at $l^2=2r_+^2$ and $\triangle_*=0$, the heat
capacity has also the same property. The divergence of thermodynamic quantities
means that the existence of critical points and the occurrence of phase 
transitions.  The first point is just the thermodynamically stable
boundary of the source rotating D3-branes \cite{Cai1}, at which the ratio 
of the critical angular velocity to the critical temperature is
\begin{equation}
\gamma_s  \equiv \left. \frac{\Omega}{T}\right |_s
       =\left. \frac{2\pi lr_+} {2r_+^2+l^2}\right
     |_{l^2=2r_+^2}=\frac{\pi}{\sqrt{2}}.
\end{equation}     
At the second point, we can also get a similar relation from $\triangle_*=0$,
\begin{equation}
\gamma _p \equiv \left.\frac{\Omega}{T} \right|_p
      =\left. \frac{2\pi lr_+}{2r_+^2+l^2}\right
     |_{2m=u^2(u^2 +l^2)}=\frac{\sqrt{2}\pi l\sqrt{\sqrt{l^4+4u^2(u^2+l^2)}
     -l^2}}{\sqrt{l^4 +4u^2(u^2+l^2)}},
\end{equation}     
which depends on the value $u/l$. For example, when $u/l=1$, one has
$\gamma_p= 2\pi/3$. The second critical point is just the extension of 
the so-called maximal 
temperature observed in \cite{Tseyt2} when the probe is in the static D3-brane
background.  Furthermore, $\triangle_*=0$ means that the probe is just
at the horizon of the rotating source. 
In the static background, the first critical point is absent. 
In the rotating D3-brane background, for fixed angular velocity and mass
scale, if $\gamma_s <\gamma_p$, the first critical point will not occur, 
that is, the whole system  is already  unstable due to the probe 
before the occurrence of the thermodynamic instability of the source; if
$\gamma_s >\gamma_p$,  the second critical point is then absent because the 
whole system is already unstable due to the thermodynamic instability of 
source before reaching the second critical temperature.

 In \cite{Tseyt2} it has been observed that at the low-temperature 
 or long-distance ($u \rightarrow \infty $) limit, the free energy of the
 probe can be expressed as 
 \begin{equation}
\label{e53}
 G_p^{\infty} =G_{N+1}-G_N-G_1,
 \end{equation}
where $G_{N+1}$, $G_N$, and $G_1$ denotes the free energies for the
$N+1$ coinciding D3-branes, $N$ coinciding D3-branes and one D3-brane,
respectively.  Hence the free energy of the probe can be explained very well
by the contribution of the massive states to the free energy of SYM theory
at strong 't Hooft coupling. In our case, taking the limit $u \rightarrow
\infty$, from (\ref{perp}) we have  
\begin{equation}
\label{e54}
G^{\infty}_{\perp} =-\frac{V_3N r_+^2 (r_+^2+l^2)}{4\pi^2 \lambda^2},
\end{equation}
while from (\ref{G_N}) 
\begin{equation}
\label{e55}
G_N =-\frac{V_3 N^2 r_+^2 (r_+^2+l^2)}{8\pi^2 \lambda^2}.
\end{equation}
As expected, the relation (\ref{e53}) holds as well in this case.

When  $\theta=\pi/2$, the free energy of the probe is
\begin{equation}
\label{e57}
G_{\parallel}=
 \frac{V_3N u^4}{2\pi^2 \lambda ^2}\left (
 \sqrt{1-\frac{2m}
     {u^4 }-\frac{\lambda  \Omega^2(u^2+l^2) }{ u^4}
      +\frac{2\sqrt{2m\lambda}l\Omega}{ u^4}
      }-1 -\frac{\sqrt{2m\lambda}l\Omega}{ u^4}
     \right).
\end{equation}
Its entropy is
 \begin{eqnarray}
S_p &=& \frac{V_3N\sqrt{2m\lambda}l^3}{\pi \lambda^2 (2r_+^4 +3l^2r_+^2 +l^4)}
   \frac{l^2+2r_+^2}{l^2-2r_+^2}
   \frac{1}{\triangle_{**}}
   \left [ \left(2m +\sqrt{2m\lambda}l\Omega
   \triangle_{**}- \sqrt{2m\lambda}l\Omega \right)\frac{l^2-2r_+^2}{l^3}
   \right. \nonumber\\
   &-& \left. \sqrt{2m\lambda}\Omega +\lambda l\Omega^2+ 
   \sqrt{2m\lambda}\Omega \triangle_{**} \right],
\end{eqnarray}
and the angular momentum 
\begin{eqnarray}
J_p &=& \frac{V_3N}{2\pi^2 \lambda^2}
       \frac{\sqrt{2m\lambda}(2r_+^2+3l^2)}{(r_+^2+l^2)(l^2-2r_+^2)}
       \frac{1}{\triangle_{**}}
       \left[ \left(2m +\sqrt{2m\lambda}l\Omega
   \triangle_{**}- \sqrt{2m\lambda}l\Omega \right)
   \frac{l(l^2-2r_+^2)}{r_+^2(2r_+^2+3l^2)}
   \right. \nonumber\\
   &+& \left. \sqrt{2m\lambda}\Omega -\lambda l\Omega^2- 
   \sqrt{2m\lambda}\Omega \triangle_{**} \right],
\end{eqnarray}
where
$$
\triangle_{**}=
  \sqrt{1-\frac{2m}
     {u^4 }-\frac{\lambda  \Omega^2(u^2+l^2) }{ u^4}
      +\frac{2\sqrt{2m\lambda}l\Omega}{ u^4}}.       
               $$
The same happens as the first case. That is, the entropy and angular momentum
of the probe is divergent at the source instability point $l^2=2r_+^2$ 
and $\triangle_{**}=0$. The remarks about the  thermodynamic instability 
in the first case are applicable here as well. However, 
from (\ref{e57}) we see that in the long-distance limit (large $u$) 
\begin{equation}
\label{e60}
G_{\parallel} \approx -\frac{V_3N}{4\pi^2\lambda ^2}\left[r_+^2(r_+^2+l^2)
      +\lambda \Omega^2 (u^2+l^2)\right].
\end{equation}
That is, when $u\rightarrow \infty$, the free energy is divergent.
Obviously  the relation (\ref{e53}) does not hold  in this case. 
For a general $\theta$, the thermodynamic properties 
of the probe is similar to the case of $\theta =\pi/2$,
 the second critical point is just determined 
by the equation (\ref{crit}). In addition, it is worth noting that 
when $\Omega =0$ in (\ref{free}), namely, for a static probe,  
 the relation (\ref{e53}) can be recovered in the limit $u\rightarrow \infty$
for any azimuth $\theta$. In this case, the second critical point is 
$2m = u^2(u^2 +l^2\cos^2\theta)$, which  corresponds
to that the probe is just at the infinite red-shift hyperplane of the 
background.

%%%%%%%%%%%%%%%%% section 4 %%%%%%%%%%%%%%%%%%%%%%%%%%%%%%%%%%%%

\section{Conclusions}

In this work we investigated the dynamics and thermodynamics of
a probe D3-brane in the rotating D3-brane background.
As is well-known, the extremal limit of the rotating D3-brane configuration
is a multicenter solution of D3-branes, which are distributed uniformly
on a disc; the excitations of rotating D3-branes are thermodynamically
stable up to a critical angular momentum density, beyond which the
heat capacity is negative. When we discussed the dynamics of the probe,
we emphasized how the probe is captured by the source and what is the effect
of the thermodynamic stability boundary on the motion of the probe. 
In the extremal background of the rotating D3-branes, the probe 
will be bounced back at some turning point if its angular momentum 
does not vanish.  Otherwise, the probe will be absorbed by the source: 
When the probe is restricted in the  disc plane, the probe will be captured 
and will locate at the edge of the disc; when the probe moves in the 
hyperplane which is orthogonal to the disc plane, the probe will be absorbed 
at the center of the disc.  In the near-extremal background,
when $\theta=0$, which means
that the probe is in the hyperplane orthogonal to the disc plane, the probe 
will be captured and locate at the horizon of the rotating D3-branes; 
in the disc plane, the probe will be bounced back at a turning point, which is
just the infinite red-shift hyperplane of the background, even when the 
probe has zero angular momentum. 
We observed that a probe with vanishing angular momentum is always captured
and can locate at the horizon if the background is the static D3-brane
configuration. The dynamic analysis of the probe brane provides an evidence
 of the assumption (which has been used in the calculations
of the Wilsonian potentials in the non-extremal backgrounds): The non-extremal
 brane ``locates'' at the horizon of the supergravity background, at least for 
the static source. In addition, nothing special happens at the 
thermodynamically stable  boundary of the rotating D3-branes from the point 
of view of the motion of the probe.

The free energy was obtained of a probe which keeps a relative static 
position to the rotating D3-branes, namely, the probe rotates along 
an orbit in the $\phi$-direction with the same angular velocity $\dot{\phi}
=\Omega$. The free energy depends on not only the Hawking temperature
of the source and the distance to the source, but also the angular
velocity and the azimuth $\theta$. From the free energy some thermodynamic
quantities, for example, the entropy, angular momentum, and heat capacity,
of the probe can be worked out.  We found that there are two critical
points, at which those thermodynamic quantities of the probe 
diverge. One point is just the thermodynamic stable boundary of the rotating
D3-branes; the other  is related to the source parameters (Hawking
temperature and angular velocity) and the position of the probe. The latter
can be regarded as the mass scale of scalar fields in the SYM theory
since the free energy of the probe can be interpreted as the contribution
of massive states to the one of SYM theory at the strong 't Hooft 
coupling. A self-consistency check of this is  the relation (\ref{e53}), which
has been found in the static source case \cite{Tseyt2}. In our case, however,
 we found that except the case of $\theta =0$, the free energy of the probe 
diverges in the long-distance limit and hence the relation (\ref{e53})
does not hold. From the viewpoint of field theory, spinning D3-branes
means  adding a chemical potential to the SYM \cite{Kraus,Gubser1},
due to it, our calculations imply that the massive states 
have  the additional contributions to the free energy of the SYM, which 
diverge as the mass scale goes to infinity. That the relation (\ref{e53})
holds in the case of $\theta =0$ reflects the fact that in fact the probe in
this position is static  and implies
that the ``angular quantum number'' of the massive states conjugate 
to the chemical potential is zero in this case. Indeed, for a static probe
in this rotating source, the relation (\ref{e53}) can be  recovered 
for arbitrary $\theta$  as we see above. In addition, the occurrence
of the second critical point means that the SYM theory has rich phase 
structures in the Higgs phase.  Finally we mention that recently it 
has been found that  the contribution of the massive states
 also diverges  as the SYM theory is on a  $S^3$ sphere \cite{Land},
in that paper the thermodynamics of a probe has been studied in the
 static D3-branes wrapping on the $S^3$ sphere.

%%%%%%%%%%%%%%%%%% Acknowledgments %%%%%%%%%%%%%%%%%%

\section* { Acknowledgments}

 The author would like to thank Prof. N. Ohta for 
carefully reading and  helpful discussions, and the referee 
for useful suggestion. This work was supported by the Japan Society
for the Promotion of Science.

\end{document}